\documentclass{article}
\usepackage{graphicx} 
\usepackage[utf8]{inputenc}
\usepackage{amsmath}
\usepackage{xcolor}
\usepackage{biblatex} 

\usepackage{subfiles} 

\usepackage{float}

\usepackage{blindtext}
\usepackage[a4paper, total={6in, 9in}]{geometry}
\usepackage[width=\textwidth, labelfont=bf]{caption}
\usepackage{enumitem}   

\usepackage{amssymb}

\makeatletter
\renewcommand{\maketitle}{\bgroup
\begin{flushleft}

  \textbf{\Large{\@title }}\par

  \small{\@author}\par
  \medskip
  
\end{flushleft}\egroup
}
\makeatother

\title{A mathematical model for the within-host (re)infection dynamics of SARS-CoV-2}
\author{
Lea Schuh$^{1,*}$, Peter V. Markov$^{1,2}$, Vladimir M. Veliov$^3$, Nikolaos I. Stilianakis$^{1,4,*}$\\
\medskip
$^{1}$European Commission, Joint Research Centre (JRC), Ispra, Italy\\
$^{2}$London School of Hygiene \& Tropical Medicine, University of London, London, UK\\
$^{3}$Institute of Statistics and Mathematical Methods in Economics, Vienna University of Technology, Vienna, Austria\\
$^{4}$Department of Biometry and Epidemiology, University of Erlangen-Nuremberg, Erlangen, Germany\\
$^*$corresponding authors \\
Emails: lea.schuh@ec.europa.eu (L.S.), nikolaos.stilianakis@ec.europa.eu (N.I.S.)}

\begin{document}

\maketitle

\section*{Abstract}
Interactions between SARS-CoV-2 and the immune system during infection are complex. However, understanding the within-host SARS-CoV-2 dynamics is of enormous importance, especially when it comes to assessing treatment options. Mathematical models have been developed to describe the within-host SARS-CoV-2 dynamics and to dissect the mechanisms underlying COVID-19 pathogenesis.
Current mathematical models focus on the acute infection phase, thereby ignoring important post-acute infection effects. We present a mathematical model, which not only describes the SARS-CoV-2 infection dynamics during the acute infection phase, but also reflects the recovery of the number of susceptible epithelial cells to an initial pre-infection homeostatic level, shows clearance of the infection within the individual, immune waning, and the formation of long-term immune response levels after infection. Moreover, the model accommodates reinfection events assuming a new virus variant with either increased infectivity or immune escape. Together, the model provides an improved reflection of the SARS-CoV-2 infection dynamics within humans, particularly important when using mathematical models to develop or optimize treatment options. \\

\noindent \textbf{Keywords:} COVID-19, infectious disease modeling, viral kinetics, mathematical modeling, immune response, target-cell-limited model, reinfection, virus variants 

\section*{Author summary}
Interactions between SARS-CoV-2 and the immune response during an infection are complex. To develop treatment options, however, it is necessary to understand how the infection spreads, is contained within, and cleared from the body. Mathematical models have been developed to describe the SARS-CoV-2 infection dynamics in humans, focusing on the early infection phase during which individuals present a measurable viral load. Thereby, they neglect long-term infection effects such as the recovery of the number of nasal cells, permanent clearance of the infection, and the formation of long-term immune response levels after infection. Here, we present a mathematical model, which not only describes the infection dynamics of SARS-CoV-2 during the initial infection phase, but also reflects the long-term post-acute infection effects. Moreover, the model allows for reinfection with a new virus variant with distinct variant properties. Together, the model provides an improved reflection of the SARS-CoV-2 infection dynamics within humans. 

\section*{Introduction}
The interactions between severe acute respiratory syndrome coronavirus type 2 (SARS-CoV-2) and the immune system during infection are complex. Viral load measurements of the upper respiratory tract provide a quantitative method to capture the within-host SARS-CoV-2 infection dynamics \cite{Woelfel2020}. Mathematical models have been developed to dissect and quantify the underlying pathogenic mechanisms, incorporating features such as viral loads, time of clinical symptom onset or infectiousness \cite{Du2022}. \\
Viral within-host dynamics are commonly described by the target-cell limited (TCL) model \cite{Canini2014}. The TCL model describes the dynamics of susceptible (target) cells, infected cells and free virus populations during the course of an infection. In the past few years, TCL-based models have been employed to also describe the within-host infection dynamics of SARS-CoV-2. Early efforts focused on quantitatively describing the measured within-host viral load dynamics in individuals \cite{Challenger2022,Wang2020} and on linking viral loads to infectiousness \cite{Ke2021,Marc2021}. These models differentiated between infectious and non-infectious virus, such as residual viral genome fragments, or by adding simple, often non-mechanistic descriptions of the human immune response \cite{Challenger2022,Ke2021,Marc2021,Wang2020}. By adding an interferon mediated refractory cell population or reduction in the infection and viral production rates, later models integrated more mechanistic formulations of the human immune response  \cite{Ke2022,Marc2023}. These models provided further insights into SARS-CoV-2 infection dynamics, revealing heterogeneity in the infectiousness levels between individuals \cite{Ke2022} and suggesting differences in viral dynamics between variants \cite{Marc2023,McCormack2023}. 
TCL-based models of SARS-CoV-2 infection dynamics have lately also been used to describe the viral load dynamics during antiviral treatment and, consequently, to obtain insights into phenomena such as viral rebound and resistance formation \cite{Perelson2023,Phan2023}.
To date, models describing the within-host SARS-CoV-2 infection dynamics typically focus on the short-term acute phase of an infection. The assumptions that these models make are often at odds with simple but highly relevant and important clinical realities, such as the re-establishment of susceptible cells to pre-infection levels, full and permanent clearance of the infection from the individual, the corresponding decline of the immune response, and the associated established long-term post-acute infection immunity levels after exposure and infection. \\
\noindent Here, we present a within-host SARS-CoV-2 model, which offers a more realistic account of the underlying infection dynamics.
While our model captures the short-term within-host infection dynamics just as well as existing models and demonstrates a good generalizability, it considerably improves earlier work by further capturing expected post-acute infection dynamics. These include the re-establishment of the number of susceptible epithelial cells to the initial pre-infection level, a full and permanent clearance of the infection from the individual, the formation of long-term post-acute infection immune response levels after infection, and the waning of immunity.
Finally, we demonstrate that the model can accommodate reinfection events, taking into account differences in infectivity and levels of immune escape of the reinfecting virus variant. 

\section*{Results}
\subsection*{Within-host SARS-CoV-2 model}
\begin{figure}[H]
\centering
\includegraphics[width=\textwidth]{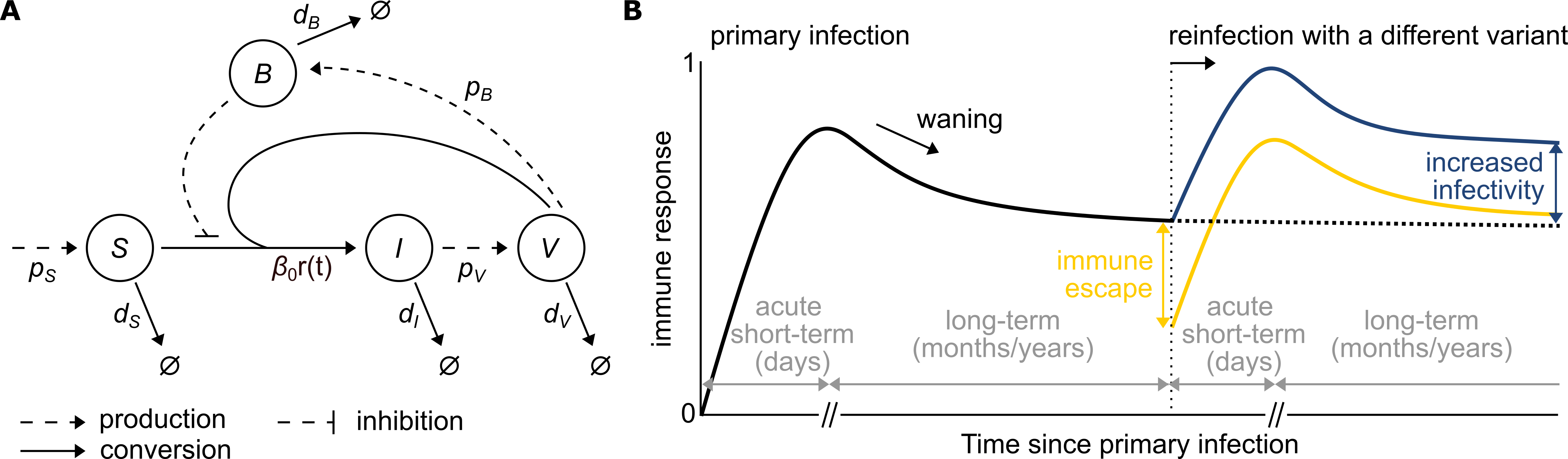}
\caption{\textbf{Schematic presentation of the within-host SARS-CoV-2 infection model and diagrammatic representation of the immune response during primary and re-infections.} (A) Susceptible cells ($S$) are produced at rate $p_S$ and become infected at rate $\beta_0 r(t)$, where $\beta_0$ is the infectivity rate, $r(t) = 1-B(t)$ is the immune response effect function and $B(t)$ is the immune response. Infected cells ($I$) in turn produce free virus ($V$) at rate $p_V$. Susceptible and infected cells die at rates $d_S$ and $d_I$, respectively, and virus is cleared at rate $d_V$. Immune response $B$ depends on the viral load $V$, is activated with rate $p_B$, and decreases with rate $d_B$. (B) Upon a primary SARS-CoV-2 infection, the acute short-term immune response detects, contains, and eliminates the virus. After viral clearance, immune protection against SARS-CoV-2 typically weakens (wanes) over time. In the case of reinfection with a different variant, the immune response is re-activated. For a different variant with immune escape (yellow), the effectiveness of the initial immune response against the different variant suffers a drop due to the variant's escape properties. For a different variant featuring higher infectivity (blue), the immune response rises to higher levels in order to successfully control the infection (blue). The dotted line represents the immune response without reinfection.} 
\label{fig:model}
\end{figure}

\noindent Our model comprises four quantities corresponding to the populations of susceptible epithelial cells ($S$) and infected epithelial cells ($I$), free virus particles ($V$), and the immune response ($B$, normalized to a range between 0 and 1) (Figure \ref{fig:model}A).  

\noindent The full system of non-linear ordinary differential equations (ODEs) for the susceptible ($S$) and infected ($I$) cells, free virus ($V$), and the immune response ($B$) is given by the following equations:
\begin{align}
\label{eq:ODEsystem} 
\begin{split}
\frac{\partial S(t)}{\partial t} &= p_S - d_S S(t) - \beta_0 r(t) S(t) V(t) \\
\frac{\partial I(t)}{\partial t} &= \beta_0 r(t) S(t) V(t) - d_I I(t) \\
\frac{\partial V(t)}{\partial t} &= p_V I(t) - d_V V(t) - \beta_0 r(t) S(t) V(t)\\
\frac{\partial B(t)}{\partial t} &= p_B (1 - B(t)) V(t) - d_B (B(t) - B_{\text{thres}})B(t),
\end{split}
\end{align}

\noindent where $r(t) = 1-B(t)$ is the immune response effect function, and $S(0) = S_0$, $I(0) = I_0$, $V(0) = V_0$, and $B(0) = 0$ are the given initial values. 
\noindent In a disease-free system, susceptible cells are in homeostasis, where cells are constantly produced at rate $p_S$ and die at a natural death rate $d_S$, where $p_S = S_0 \times d_S$. Once a free virus is introduced, the system is perturbed as follows: 
When in contact with free virus, susceptible cells get infected according to an overall infectivity rate $\beta_0 r(t)$, which is the product of the infectivity rate constant $\beta_0$ and immune response effect function $r(t)$. The immune response effect function indicates the remaining proportion of successful infections of susceptible cells despite the presence of the immune response. The term 
\begin{equation*}
    -\beta_0 r(t) S(t) V(t) = -\beta_0 (1 - B(t)) S(t) V(t) = \underbrace{-\beta_0 S(t) V(t)}_{\text{(i)}}  + \underbrace{\beta_0 B(t) S(t) V(t)}_{\text{(ii)}}
\end{equation*} 
of the first equation in (\ref{eq:ODEsystem}) is to be interpreted as (i) the removal of susceptible cells due to infection assuming no activity of the immune response ($B(t)$ = 0) and (ii) the immediate re-introduction of susceptible cells involved in unsuccessful infections inhibited by the anti-SARS-CoV-2 activity of the immune response.
The successfully infected cells enter the infected cell population and are eliminated from that population with virus-induced death rate $d_I$. Due to the non-lytic property of SARS-CoV-2, infected cells constantly produce and release free virus at rate $p_V$, which is cleared at rate $d_V$ \cite{Ghosh2020}. Free virus leading to successfully infected cells are lost from this population according to the overall infectivity rate $\beta_0 r(t)$. 
A SARS-CoV-2 infection activates the immune response (Figure \ref{fig:model}B). First, the innate immune response detects the viral infection and limits its spreading within the host. The innate immune response also triggers the adaptive immune mechanisms, which ultimately clear the virus and form a lasting, long-term SARS-CoV-2-specific immune memory \cite{Cox2020,Rodda2021,Sette2021}. This ensures an enhanced response to future infections of the same virus. The term immune response $B$ in our model represents the combined effects of innate and adaptive immunity on the ability to control the virus. For a primary infection, the individual was assumed to not have been exposed to SARS-CoV-2 before, and hence, the immune response is initially described by a naive pre-infection state with $B(0) = 0$. Upon infection, $B$ increases proportional to the viral load at rate $p_B$ and is limited by the upper bound of $1$, such that $B(t) \in [0,1]$. The activation of the immune response is assumed to lead to a reduced overall infectivity and hence, to fewer successful infections of susceptible cells \cite{Ke2021,Marc2023}. Long-term, we assumed that the immune response wanes to a low level of immunity after infection. To counteract a repeated spreading of the same virus within an individual, we defined a minimal immune threshold $B_{\text{thres}}$ against which the immune response $B$ converges to in the long-term with rate $d_B$, while assuring a permanent clearance of the infection. The immune threshold $B_{\text{thres}}$ is given by:  
\begin{equation}
\label{eq:Bthres1} 
B_{\text{thres}} = 1- \frac{d_I d_V}{\beta_0 S_0 (p_V-d_I)}.
\end{equation}
How $B_{\text{thres}}$ is determined is described in detail in the section Methods. To keep the immune response inactive during the initial virus-free state and allow for its activation upon contact with SARS-CoV-2 only, we multiplied $d_B(B(t)-B_{\text{thres}})$ by $B(t)$ in the last equation of (\ref{eq:ODEsystem}). In summary, the overall infectivity rate is
\[   
\beta_0 r(t) = 
     \begin{cases}
       \beta_0, &\quad\text{if there is no immune response}\\
       0, &\quad\text{if the immune response reaches maximal capacity}\\
       \beta_0 (1-B_{\text{thres}}), &\quad\text{if the individual attained long-term immune response levels}\\
       \beta_0 r(t), &\quad\text{otherwise.}\\ 
     \end{cases}
\]
After a successful infection and clearance of the virus, the susceptible cells return to the initial pre-infection level. Steady state and stability analyses demonstrate that this is the only post-disease steady state (see Methods). Overall, the model is described by four quantities, $S$, $I$, $V$, and $B$, and eight parameters, $p_S$, $d_S$, $\beta_0$, $d_I$, $p_V$, $d_V$, $p_B$, and $d_B$, where all parameters are restricted to positive values for biological interpretability. 

\subsection*{Infection dynamics}
\begin{figure}[H]
\centering
\includegraphics[width=\textwidth]{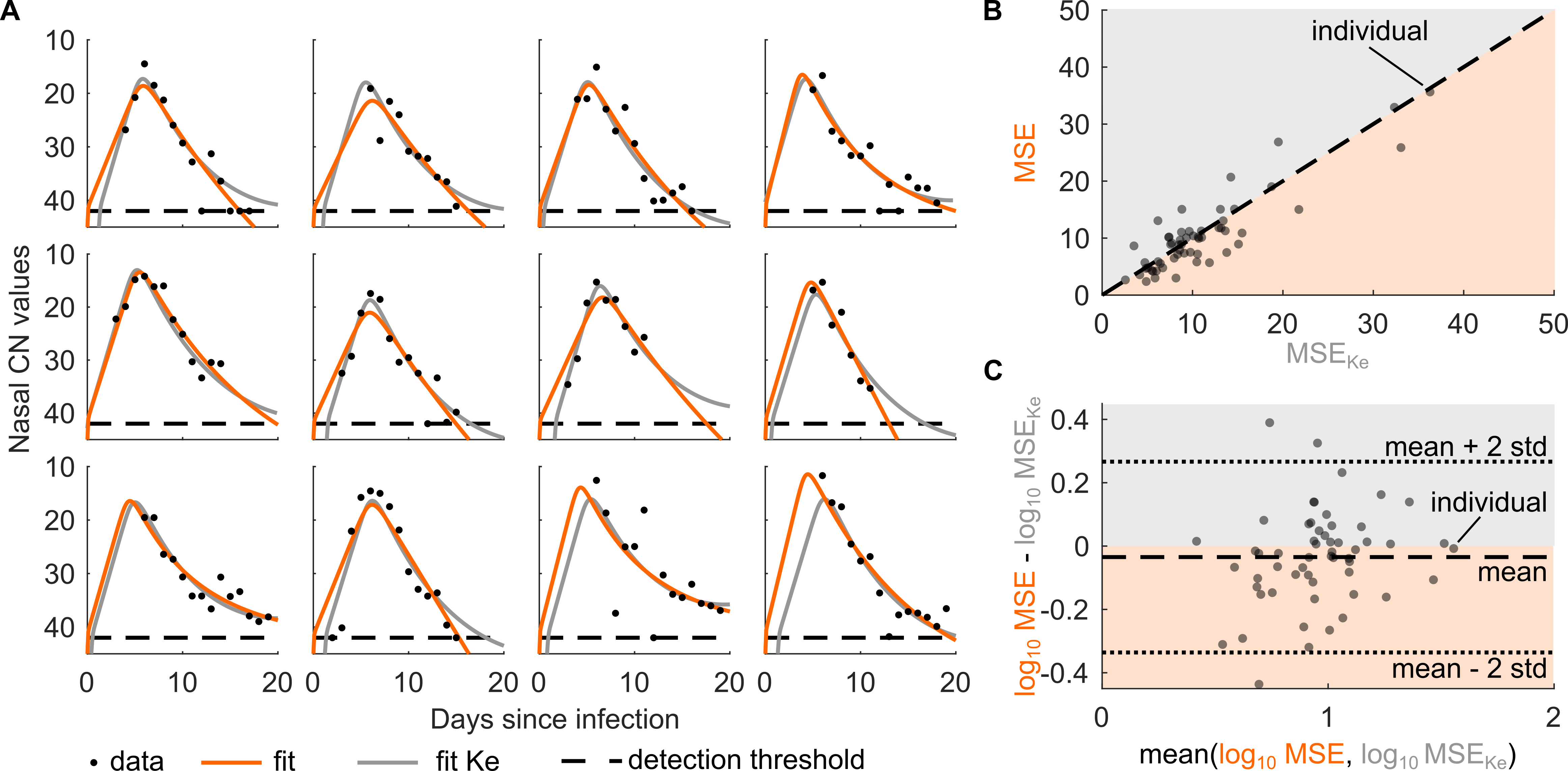}
\caption{\textbf{Model captures the individual-specific viral load dynamics of a primary SARS-CoV-2 infection.} (A) Model fits of our model (orange line) and Ke model (gray line) to nasal CN values (black dots, measured by Ke et al.) of 12 randomly selected individuals. The dashed line represents the detection threshold of the RT-qPCR method used by Ke et al. to determine viral presence in the nasal swab sample and is set to CN = 42. Dots on the dashed line denote measurements below the detection threshold. (B) Comparison scatter plot of the mean squared errors (MSEs) of our model fits to fits from Ke et al. (MSE$_{\text{Ke}}$) for all 56 individuals. The dashed line indicates the diagonal representing equal MSEs between model fits. All points above the diagonal represent individuals for which the Ke model fits lead to smaller MSEs than our model (gray shaded area), while all points below the diagonal represent individuals, for which our model fits lead to smaller MSEs than the Ke model (orange shaded area). (C) Comparison of $\text{log}_{\text{10}}$ MSEs of model fits to all 56 individuals according to the Bland-Altman method. The mean of the $\text{log}_{\text{10}}$ MSE differences between our model fits and those from Ke et al. is -0.035 (black dashed line) and the mean of the $\text{log}_{\text{10}}$ MSE differences $\pm$2 $\times$ standard deviations (stds) are -0.34 and 0.27, respectively (black dotted lines). All points above zero represent individuals for which the Ke model fits lead to smaller $\text{log}_{\text{10}}$ MSEs than our model (gray shaded area), while all points below zero represent individuals, for which our model fits lead to smaller $\text{log}_{\text{10}}$ MSEs than the Ke model (orange shaded area).} 
\label{fig:fits}
\end{figure}

\noindent To quantitatively describe the dynamics of a SARS-CoV-2 infection in a human, we fitted the model to 56 individual nasal viral load samples from Ke et al. Ke et al. used reverse transcriptase polymerase chain reaction (RT-PCR) to detect SARS-CoV-2 by amplifying viral RNA in a series of repeated cycles. The number of reaction cycles needed to reach a certain cycle threshold value determines the positivity of a test and is defined as the cycle number (CN). The lower the CN value the greater the amount of viral RNA present in the original sample. To describe the transformed CN values using the model, we formally redefined population $V$ as the virus in the nasal swab sample and, similarly, $p_V$ as the viral production rate times the sampled virus \cite{Ke2021}. To account for the heterogeneity between individuals, we estimated three of the eight model rate constants, namely the viral production rate constant $p_V$, the activation rate constant of the immune response $p_B$, and the waning rate constant $d_B$, at an individual-specific level. All other rate constants were assumed to be fixed and the same across individuals. We observed a good fit of our model to the individual-specific CN values with similar dynamics as predicted by the model of Ke et al. (Figure \ref{fig:fits}A). The main differences between model fits are in the predicted initial infection dynamics prior to peak viral loads. 

\noindent From the estimated individual-specific viral production rate constants $p_V$ and the fixed death rate constant of an infected cell $d_I$, we calculated an estimated median number of $75$ viral particles (interquartile range [66, 98]) being produced during the life span of an infected cell. The average produced number of viral particles per infected cell and individual is given by $\frac{p_V}{d_I}$. 
Furthermore, we qualitatively compared the fits of our model to those of Ke et al. for all 56 individuals, using the mean squared errors (MSEs) of the respective model fits (Figure \ref{fig:fits}B). The MSEs are closely scattered around the diagonal line, suggesting a good general agreement between the two models across individuals. For 33 of the 56 individuals, our model fits led to lower MSEs than those of Ke et al. (dots in orange area of Figure \ref{fig:fits}B). Furthermore, we considered the Bland-Altman method for statistically assessing the agreement between the model fits \cite{Bland1986}. Considering the log$_{10}$ MSEs, we found a negligible bias (log$_{10}$ MSE - log$_{10}$ MSE$_{\text{Ke}}$) and identified that in $95\%$ of the individuals the MSEs of Ke et al. are between $0.46$ and $1.85$ times the MSEs of our model (Figure \ref{fig:fits}C). Overall, the two models concur well in capturing the SARS-CoV-2 infection dynamics of nasal viral load samples in humans.

\subsection*{Numerical results}
\begin{figure}[H]
\centering
\includegraphics[width=0.9\textwidth]{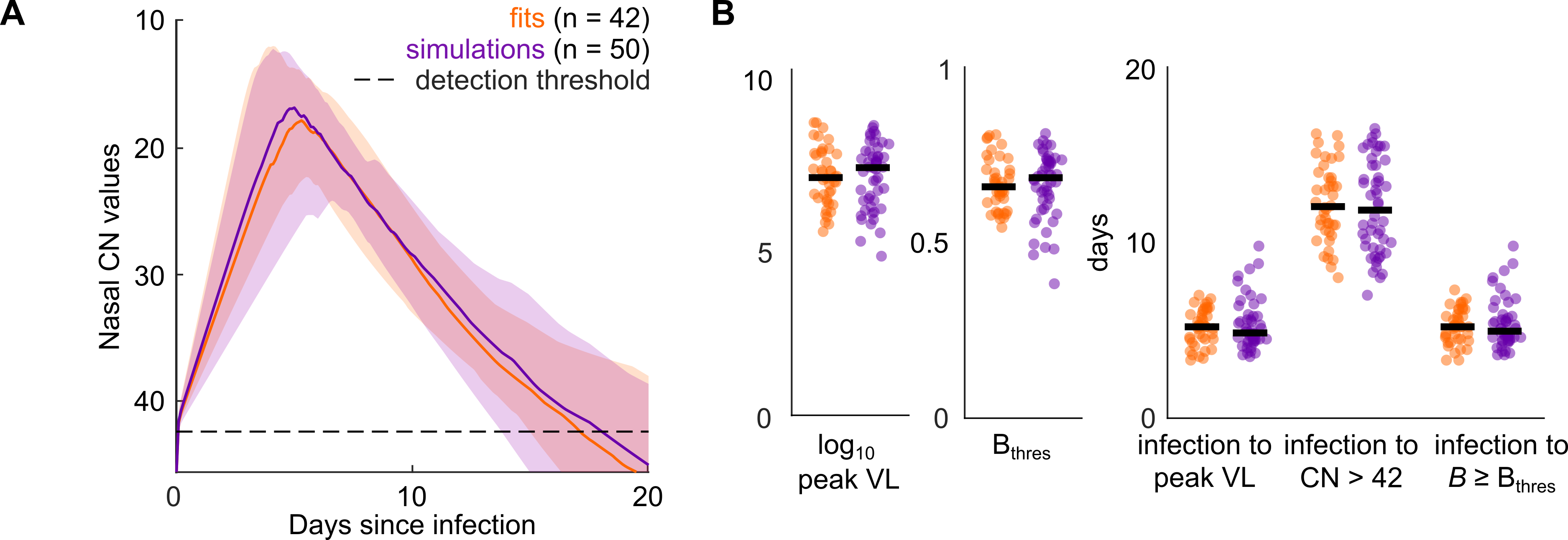}
\caption{\textbf{Model reproduces key features of SARS-CoV-2 infection dynamics.} 
(A) Median, 10$^{\text{th}}$ to 90$^{\text{th}}$ percentiles of 42 fits of our model to nasal CN values (Ke et al., orange) and 50 simulated infection dynamics (purple) using our within-host model. The dashed line represents the detection threshold of the RT-qPCR method used by Ke et al. to determine viral presence in the nasal swab sample and is set to CN = 42. (B) Distributions of the five key features describing the underlying infection dynamics: (i) log$_{10}$ peak viral loads (VL), (ii) $B_{\text{thres}}$-values and durations from (iii) infection to peak VL, (iv) peak VL to undetectable VL, defined as a CN value > 42, and (v) infection to $B > B_{\text{thres}}$ across the fits (orange) and simulations (purple). The distributions of key features across fitted and simulated infection dynamics are statistically comparable with \textit{p}-values 0.37, 0.51, 0.42, 0.52, and 0.45, respectively (two-sample Kolmogorov-Smirnov test correcting for five-fold testing according to the Bonferroni correction). The scatter is consistent between fits and simulations across sub-panels. The median values per feature are highlighted by the black bars.}
\label{fig:sim_study}
\end{figure}

To test the generalizability of the model, we performed a simulation study. Similar to the model fitting in the previous section, we assumed the rate constants $p_V$, $p_B$, and $d_B$ to be individual-specific and to capture the heterogeneity underlying the SARS-CoV-2 infection dynamics. All other rate constants were assumed to be fixed and to be the same across individuals. We drew 50 rate constant triplets out of the multivariate distribution we received from the estimated individual-specific rate constants in the previous section. Using these sets of rate constants and our within-host model, we simulated 50 SARS-CoV-2 infection dynamics (Figure \ref{fig:sim_study}A).

\noindent From all 42 fits and 50 simulations, we computed five key features describing the underlying infection dynamics, namely (i) the log$_{10}$ peak viral load (VL), (ii) $B_{\text{thres}}$, (iii) the number of days from infection to peak VL, (iv) the number of days from peak VL to undetectable VL (defined as a CN value > 42), and (v) the number of days from infection to $B \geq B_{\text{thres}}$. We found the distributions of each of these five key features to be statistically comparable, that is we cannot reject the null hypothesis that the medians of both groups are equal, between the fitted and simulated infection dynamics with \textit{p}-values 0.37, 0.51, 0.42, 0.52, and 0.45, respectively (Figure \ref{fig:sim_study}B). 
The median values for each of the key features for our fitted and simulated infection dynamics are shown in Table~\ref{tab:median_val} alongside those reported by earlier studies.

\begin{table}[h]
\centering
\begin{tabular}{c l c c c c}
\hline
 & & \multicolumn{3}{c}{Median value} & \\
 & Key features of infection dynamics & Fits & Simulations &  Other studies & Ref. \\
 \hline
(i) & log$_{10}$ peak VL  & 7.2 & 7.5 & 4 - 9 & \cite{Challenger2022, Killingley2022,Woelfel2020}\\
(ii) & $B_{\text{thres}}$ &  0.66 & 0.68 & - & - \\
(iii) & Days from infection to peak VL & 5.2 & 4.9 & 4 - 6 & \cite{Killingley2022} \\
(iv) & Days from peak to undetectable VL & 12 & 12 & 8 - 11 & \cite{Killingley2022,Woelfel2020} \\
(v) & Days from infection to $B \geq B_{\text{thres}}$ & 5.2 & 5.0 & - & - \\
\end{tabular} 
\caption{\textbf{Median values of key features of infection dynamics: output from our model (fitted and simulated) alongside corresponding values reported by earlier studies.}}
\label{tab:median_val}
\end{table}

\noindent Our model and simulation output agree well with those reported by others. Together, the model reliably generates simulated individual-specific SARS-CoV-2 infection dynamics.

\subsection*{Post-acute infection and reinfection dynamics}
\begin{figure}[H]
\centering
\includegraphics[width=\textwidth]{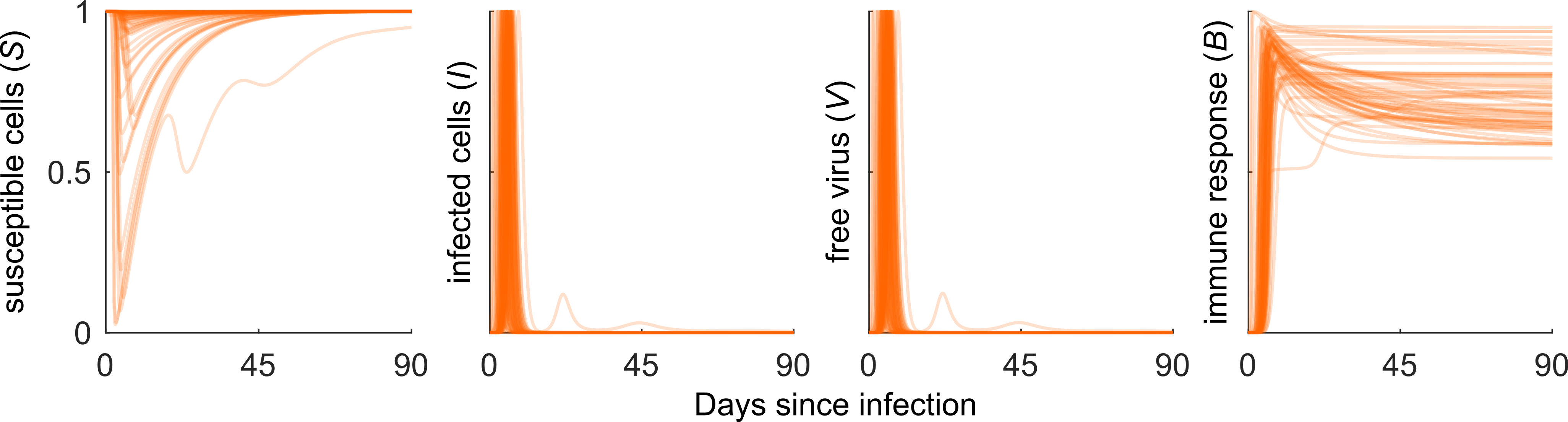}
\caption{\textbf{SARS-CoV-2 infection dynamics following acute infection.} SARS-CoV-2 infection dynamics for the populations of susceptible and infected cells $S$ and $I$, respectively, free virus $V$, and the immune response $B$ for all 56 fitted individuals following acute infection. The susceptible cells are normalized per individual by the initial number of susceptible cells $S_0$, while the number of infected cells and free virus particles are normalized by their maximal values during infection.}
\label{fig:LTdynamics}
\end{figure}

To verify the model performance for biological realism of its output, we further considered our model fits of the nasal viral load data (Ke et al. \cite{Ke2022}) for a period of 90 days post infection (Figure \ref{fig:LTdynamics}).
\noindent For all but one individual, the susceptible cells $S$ re-establish their initial pre-infection levels after infection, which is denoted by $S$ = 1 at 90 days since infection (Figure \ref{fig:LTdynamics}, left panel). Moreover, the infection is permanently and fully cleared within all individuals, given by $I$ = 0 and $V$ = 0 after 90 days since infection (Figure \ref{fig:LTdynamics}, middle panels). Finally, the model fits of all individuals account well for immune waning and the establishment of long-term post-acute infection immune response levels at $B_{\text{thres}}$ after infection (Figure \ref{fig:LTdynamics}, right panel, individual-specific $B_{\text{thres}}$-values not shown). Through its intrinsic properties, our within-host SARS-CoV-2 model captures effectively the expected post-acute infection dynamics with respect to the different model compartments. This is an improvement with respect to earlier models, which provide anomalous predicted infection dynamics when extended to a long-term post-acute infection period.\\

\begin{figure}[H]
\centering
\includegraphics[width=0.9\textwidth]{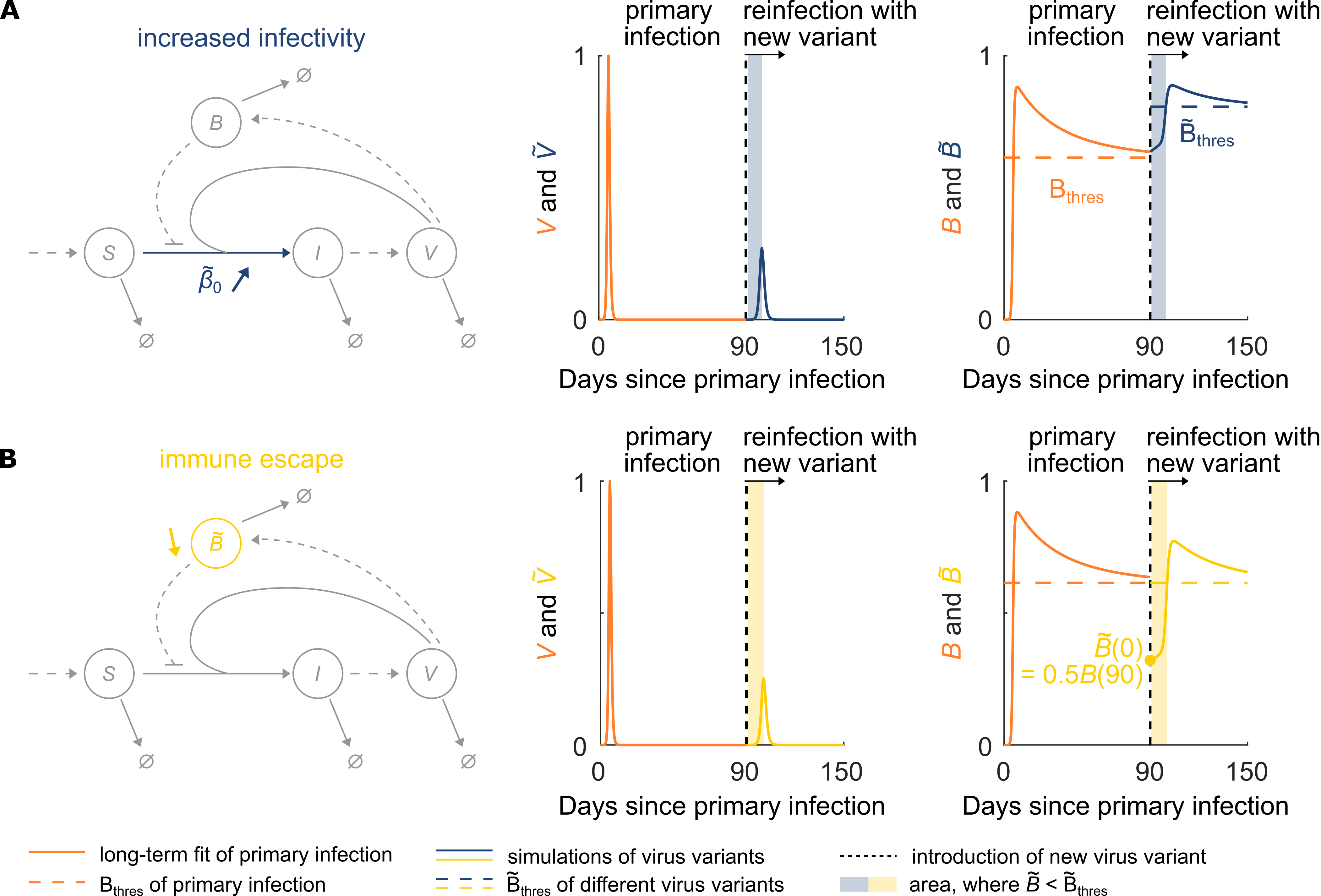}
\caption{\textbf{Model fits and simulation of reinfection with virus variants with two different changes in properties: increased infectivity and immune escape.} (A) Schematic presentation of the adapted model parametrization accounting for a different virus variant with increased infectivity rate constant $\tilde{\beta}_0$ (here, $\tilde{\beta}_0 = 2 \beta_0$). Model fit of the viral load $V$ and immune response $B$ during and post-acute infection of a representative individual is shown in orange. The vertical dotted black line indicates the day of contact with the respective virus variant. The continued simulation demonstrates an increase in the viral load of the variant $\tilde{V}$, and hence, reinfection (blue line). The variant-specific immune response dynamics of $\tilde{B}$ and corresponding variant-specific immune threshold $\tilde{B}_{\text{thres}}$ are shown in the right panel (blue line and blue dashed line, respectively). The blue shaded area corresponds to the duration for which $\tilde{B} < \tilde{B}_{\text{thres}}$ and indicates the period at which the new virus variant with increased infectivity can spread within the host. (B) Schematic presentation of the adapted model parameterization accounting for a different virus variant with immune escape reflected by a decreased initial variant-specific immune response $\tilde{B}$ (here, $\tilde{B}(0) = 0.5 B(90)$). Like in (A), the model fit of viral load $V$ and immune response $B$ during and post-acute infection of a representative individual is shown in orange. The vertical dotted black line indicates the day of contact with the respective virus variant. The continued simulation demonstrates reinfection (yellow line). The variant-specific immune response dynamics of $\tilde{B}$ and corresponding variant-specific immune threshold $\tilde{B}_{\text{thres}}$ are shown in the right panel (yellow line and dashed line, respectively). The yellow shaded area corresponds to the duration for which $\tilde{B} < \tilde{B}_{\text{thres}}$ and indicates the period at which the new virus variant with immune escape can spread within the host.}
\label{fig:reinfection}
\end{figure}

\noindent According to the USA Centers of Disease Control and Prevention, a reinfection is a positive detection of SARS-CoV-2 at least $90$ days after a previous infection \cite{CDC2020}. At present, reinfections typically involve a different SARS-CoV-2 variant and hence, potentially different infection dynamics due to differences in both the reinfecting variant or the variant-specific immune response \cite{Nguyen2022}. 
In our model parameterization, we accounted for both of these factors. We explored a scenario, where the new variant features an increased infectivity, which is reflected by an increased infectivity rate constant $\beta_0$ in our model  (Figure \ref{fig:reinfection}A, left panel). Furthermore, we explored a scenario, where the new variant features immune escape. In our model, immune escape is reflected  by an initial drop in the variant specific immune response $B$ (Figure \ref{fig:reinfection}B, left panel).
\noindent We used the previously simulated post-acute primary infection dynamics of a randomly selected representative individual from the Ke et al. data set. Furthermore, we assumed the individual to have been in contact with a different variant with either increased infectivity or immune escape, leading in each of these scenarios to a single successfully infected cell at 90 days since primary infection. Using the adapted variant-specific model parameterizations, that is an increased infectivity rate constant of $\tilde{\beta}_0 = 2 \beta_0$ or a decreased initial immune response of $\tilde{B}(0) = 0.5 B(90)$, where $\tilde{\beta}_0$ the variant-specific infectivity rate constant and $\tilde{B}(t)$  the variant-specific immune response, we simulated the viral loads and immune responses for both variants for another 60 days after reinfection (Figure \ref{fig:reinfection}A and B, middle and left panels).   
Our corresponding simulation results show a detectable increase in the viral loads due to reinfection for both variants. As a result of pre-existing immunity at the time of reinfection and given the parameterizations of $\tilde{\beta}_0$ and $\tilde{B}(0)$, the peak viral loads remain lower during reinfection than observed during primary infection for both variants. Overall, the model accommodates well for reinfection events and differentiates between reinfection modes involving distinct variant properties.

\section*{Discussion}
We developed a mathematical model that describes the within-host SARS-CoV-2 infection and reinfection dynamics. Our model efficiently captures the acute short-term viral infection dynamics of the virus, and importantly, it further successfully accommodates biological phenomena surrounding the period following acute infection. By including viral loads and immune response dynamics during reinfection events by SARS-CoV-2 variants with increased infectivity or immune escape properties, our model accounts for aspects of the within-host infection process that go beyond current approaches. \\ 
We fitted the model to data from a previous clinical study that measured viral loads of infected individuals and found the model to describe the acute short-term infection dynamics equally well as the model fits of Ke et al. (Figure \ref{fig:fits}). 
We fixed all model parameters for which there are reliable value estimates available from the literature and only estimated the three rate constants for which literature yields a very wide range of possible values or no values at all. These three rate constants are the viral production rate constant and the two rates determining the dynamics of the immune response: the activation and the waning rate constants. Numerical results showed that the fits of our model to the individual viral load are sensitive to these three model parameters. 
Small differences in the initial infection dynamics prior to peak viral loads may be the result of our model assuming that peak viral load occurs at 6 days post infection, while Ke et al. estimated the duration from infection to peak viral load. As there is only little data collected before the peak viral load, it is difficult to comment on these discrepancies between models. Using our model fits, we estimated that a single infected cell produces on average $75$ viral particles in total (interquartile range [66, 98]) during its life span. The number of infectious viral particles cells produce is an important property in studying intra-host viral dynamics. Our value is in the range of those reported by earlier studies: from 10 to 100 infectious particles produced by a single infected cell \cite{Goncalves2020,Sender2021}. \\
Due to the time-dependent description of the immune response and the introduction of an immune threshold, $B_{\text{thres}}$, which is the minimal immune response required to counter a repeated spreading of the same virus variant within the individual, our approach provides insights into additional features regarding the course of infection within an individual. In the post-acute infection phase, the model recapitulates the restoration of the susceptible cell numbers to pre-infection homeostatic levels and accounts for the complete clearance of the infection, for which the immune threshold is critical (Figure \ref{fig:LTdynamics}). After an acute infection, the immune response $B$ converges to the immune threshold and, hence, maintains at long-term post-acute infection levels. The immune threshold in our model may reflect the persistence of the underlying immune response across sequential SARS-CoV-2 infections as observed by Kissler et al. \cite{Kissler2023}.
For the definition of $B_{\text{thres}}$, we determined the basic reproduction number $R_0$, when no immune response is present, and an effective reproduction number $R$ that is immune response-dependent and changes over time (Figure~\ref{fig:R0}). Stability analysis of the model allows for a lucid description of the dynamic behaviour of the populations involved in the infection process and their steady states.
Together, through its intrinsic properties, our within-host SARS-CoV-2 model captures effectively the expected long-term post-acute infection dynamics with respect to the different model populations. This is an improvement on earlier models, which when extended to the post-acute infection period provide anomalous predicted infection dynamics. \\
Our model further incorporates reinfection events by SARS-CoV-2 variants with different properties affecting within-host viral dynamics (Figure \ref{fig:reinfection}). Introducing a new virus variant with increased infectivity or immune escape perturbs the convergence of $B$ to the immune threshold by raising the threshold or by lowering the effective immunity, respectively. As a result, the immune response against the new virus variant at the time of secondary infection is lower than the new immune threshold (Figure~\ref{fig:LTdynamics}). According to the formal mathematical definition of the immune threshold of the virus variant  $\tilde{B}_{\text{thres}}$ in our model, the virus variant spreads if $\tilde{B} < \tilde{B}_{\text{thres}}$ (Figure \ref{fig:R0}). Only when $\tilde{B}$ reaches a value greater than $\tilde{B}_{\text{thres}}$ is the spread of the new virus variant contained by the new updated immune response $\tilde{B}$. For both kinds of new virus variants, this feature allows the model to account for the initiation of the reinfection. That ability of our model to accommodate reinfection events and differentiate between reinfection modes involving distinct variant properties allows for its applications in clinical and public health scenarios, providing a more realistic description of key outcomes. \\ 
\noindent This study has several limitations. 
In contrast to other existing models, our within-host SARS-CoV-2 model does not account for the time delay between a cell being infected and it becoming infectious, i.e., actively producing virus, termed eclipse phase (Figure \ref{fig:model}A). Considering that the acute infection is on average 20 days \cite{Du2020}, an eclipse period of about 6 hours \cite{Ke2021} is short in comparison. Hence, we do not expect the omission of the eclipse phase to change the results qualitatively.
Moreover, the immune response in the model is summarized by the dynamic variable $B$, which regulates the overall infectivity. 
There is a large variation on how modelling approaches incorporate the immune response and its effects on the viral dynamics of SARS-CoV-2 throughout literature \cite{Challenger2022,Ke2022,Ke2021,Marc2023}. Our model uses an approach, which properly captures more of the dynamic features of the immune response, such as waning immunity, while at the same time describing the short-term viral load dynamics as clinically observed (Figure \ref{fig:fits}A). Therefore, we believe our model assumptions to be justified.
Furthermore, our aim was to develop a within-host SARS-CoV-2 model, which not only describes the acute infection phase, but also the post-acute infection dynamics and reinfection. We therefore assumed a simple optimization approach for describing the individual viral load dynamics and performed a qualitative comparison between our model fits and the original fits of Ke et al. (Figure \ref{fig:fits}). 
Finally, it must be noted that the possibility of reinfection with the same virus variant is not accounted for in the model. While, in theory this is possible as a result of waning immunity, in reality the slow nature of waning combined with the rapid evolution of SARS-CoV-2 makes is unlikely for the variant causing the primary infection to be around by the time immunity is low enough to allow reinfection with it. \\
\noindent Although developed for the infection of cells in the upper respiratory tract the model can be easily adapted to account for the lower respiratory tract and extended to assess the effects of treatment options and drug resistance development. \\
Overall, the model presents an advance in the description of the SARS-CoV-2 infection dynamics within humans capturing the full dynamic infection process from initial infection to the clearance of the virus to reinfection and the corresponding short-term and persisting immune response across multiple infections. This modelling approach should be of interest for clinical use when quantitatively describing the within-host SARS-CoV-2 infection and developing treatment options seeking optimal treatment design. 

\section*{Methods}
\subsection*{Determining immune threshold $B_{\text{thres}}$}
\begin{figure}[H]
\centering
\includegraphics[width=0.55\textwidth]{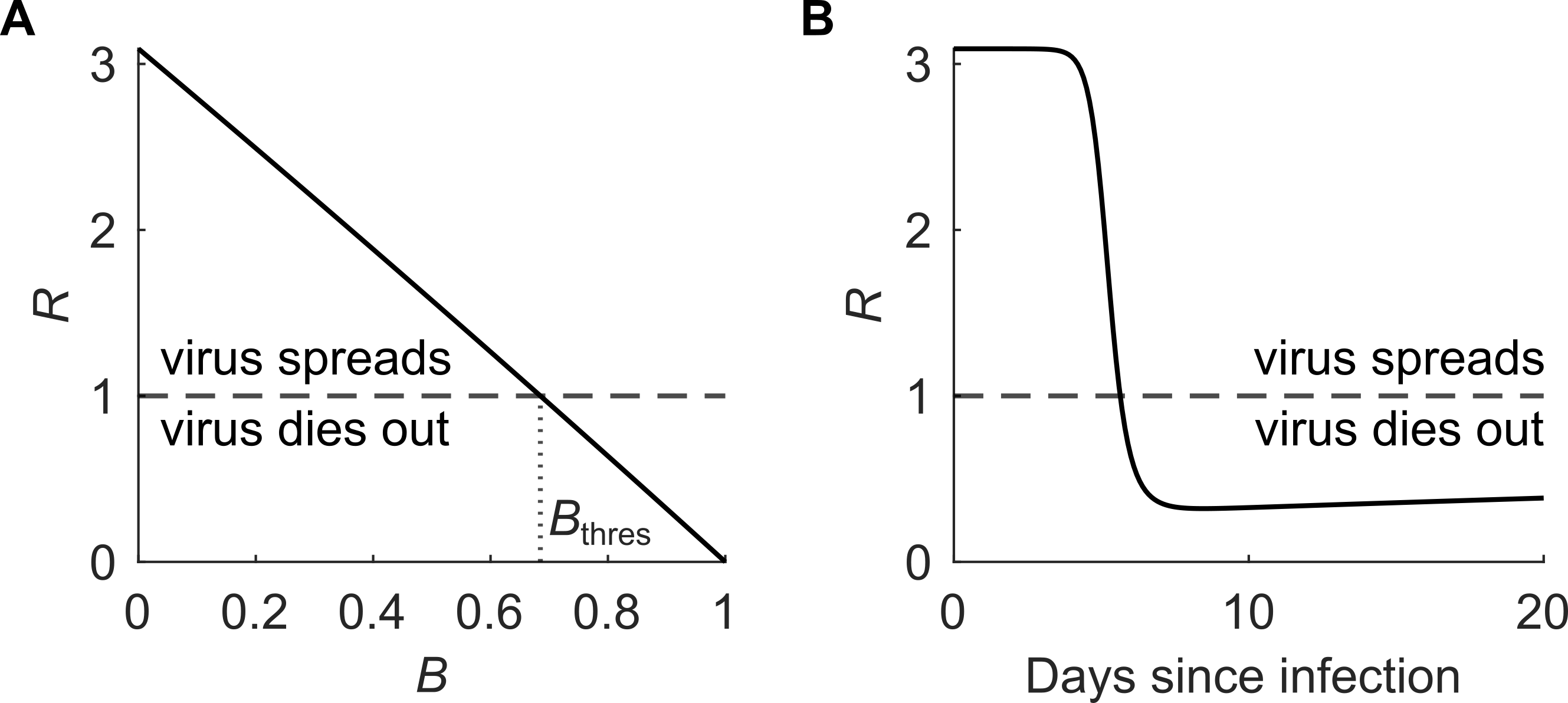}
\caption{\textbf{Changes of the reproduction number $R$ over the immune response $B$ (A) and over the course of an infection (B)}. The parameter values are taken from Table \ref{tab:nasal_saliva_par}, with $p_V = 200$, $p_B = 10^{-8}$, and $d_B = 10^{-2}$. The horizontal dashed line denotes $R = 1$ and the vertical dotted line denotes the corresponding value of $B$, defined as the immune threshold $B_{\text{thres}}$.} 
\label{fig:R0}
\end{figure}

We defined immune threshold $B_{\text{thres}}$ as the minimal immune response required to counter a repeated spreading of the same virus variant within an individual. To determine $B_{\text{thres}}$, we first derived the basic reproduction number $R_0$, which signifies the number of secondary virus particles generated by the infection of a single virus when introduced into a fully susceptible cell population, hence, $S = S_0$. If $R_0 > 1$, more virus particles are generated each generation leading to an increase of the viral population. If $R_0 < 1$, less virus particles are generated each generation letting the infection in the individual run out over time. To calculate $R_0$, we first determined the Jacobian of the infected subsystem including compartments $I$ and $V$:

\begin{align*}
J_{\text{sub}} &= 
\begin{pmatrix}
-d_I & \beta_0rS_0 \\
p_V & -d_V-\beta_0rS_0
\end{pmatrix} \\
&= 
\underbrace{
\begin{pmatrix}
0 & \beta_0rS_0 \\
0 & 0
\end{pmatrix}}_{:= T}
+ 
\underbrace{
\begin{pmatrix}
-d_I & 0 \\
p_V & -d_V-\beta_0rS_0
\end{pmatrix}}_{:= \Sigma},
\end{align*}

\noindent where $T$ corresponds to the transmission and $\Sigma$ to the transition matrix \cite{Diekmann2010}. The next-generation matrix NGM is then given by
\begin{align*}
    \text{NGM} &= -T \Sigma^{-1} \\
    &= \frac{1}{d_I(d_V+\beta_0rS_0)} \begin{pmatrix}
p_V\beta_0rS_0 & d_I\beta_0rS_0 \\
0 & 0
\end{pmatrix},
\end{align*}
whose dominant eigenvalue is the reproduction number $R$, where $R$ is dependent on $B$ and hence time $t$ (Figure \ref{fig:R0}):
\begin{equation}
    R = \frac{p_V\beta_0rS_0}{d_I(d_V+\beta_0rS_0)} =  \frac{p_V\beta_0(1-B)S_0}{d_I(d_V+\beta_0(1-B)S_0)}.
\end{equation}

\noindent The reproduction number $R$ can be split into  
\begin{equation*}
    R = \frac{p_V\beta_0rS_0}{d_I(d_V+\beta_0rS_0)} = \underbrace{\frac{p_V}{d_I}}_{\text{(i)}} \underbrace{\frac{\beta_0rS_0}{(d_V+\beta_0rS_0)}}_{\text{(ii)}},
\end{equation*}
where (i) represents the number of viral particles produced during the life span of an infected cell and (ii) denotes the number of susceptible cells infected during the life span of a free viral particle. At the time of infection, the immune response is naive, such that the initial reproduction number $R_0$ without immune response ($B = 0$) is given by:
\begin{equation*}
    R_0 = \frac{p_V \beta_0 S_0}{d_I(d_V+\beta_0 S_0)}.
\end{equation*}
The value $B_{\text{thres}}$ is then determined so that $R = 1$ for $B = B_{\text{thres}}$. 
In the case $p_V \leq d_I$, it is easy to show that $R <1$ for any $B \in [0,1]$. This means that infection is not possible in this case, therefore we futher assume that $p_V > d_I$.
Assuming the susceptible cells $S$ reach their homeostatic level $S_0$ at the end of an infection, $B_{\text{thres}}$ in ODE system (\ref{eq:ODEsystem}) is given by
\begin{equation}
\label{eq:Bthres} 
B_{\text{thres}} = 1- \frac{d_I d_V}{\beta_0 S_0 (p_V-d_I)}.
\end{equation}
We assume $B_{\text{thres}} > 0$, to ensure the possibility of infection within the individual. 

\subsection*{Steady states}
The steady states are defined as the states for which the right-hand sides of the ODE system (\ref{eq:ODEsystem}) are all equal to zero. We directly see that the pre-infection steady state is given by 
\begin{center}
$(S^*,I^*,V^*,B^*) = (S_0, 0, 0, 0)$. 
\end{center}
To identify the stability of the disease-free steady state, we first considered the Jacobian of the full ODE system described by (\ref{eq:ODEsystem}),
\begin{equation}
J = 
\begin{pmatrix}
-d_S-\beta_0(1-B)V & 0 & -\beta_0(1-B)S & \beta_0SV \\
\beta_0(1-B)V & -d_I & \beta_0(1-B)S & -\beta_0SV \\
-\beta_0(1-B)V & p_V & -d_V-\beta_0(1-B)S & \beta_0SV \\
0 & 0 & p_B(1-B) & -p_BV -2 d_BB + d_B B_{\text{thres}},
\end{pmatrix}
\end{equation}
and determined the Jacobian at the disease-free steady state:
\begin{equation*}
J|_{(S^*,I^*,V^*,B^*)} = 
\begin{pmatrix}
-d_S & 0 & -\beta_0S_0 & 0 \\
0 & -d_I & \beta_0S_0 & 0 \\
0 & p_V & -d_V-\beta_0S_0 & 0 \\
0 & 0 & p_B & d_B B_{\text{thres}}
\end{pmatrix}.
\end{equation*}
We then identified the eigenvalues from
\begin{align*}
\text{det}(J|_{(S^*,I^*,V^*,B^*)} - \lambda I) = &(-d_S - \lambda) (-d_I - \lambda) (-d_V -\beta_0 S_0 - \lambda) (d_B B_{\text{thres}} - \lambda)\\
&-(-d_S - \lambda) p_V \beta_0 S_0 (d_B B_{\text{thres}} - \lambda) \\
= &0,
\end{align*}
with $\lambda_1  = -d_S < 0$ and $\lambda_2 = d_B B_{\text{thres}} > 0$. From the remaining quadratic equation  
\begin{equation*}
    \lambda^2 + \underbrace{(d_I + d_V + \beta_0 S_0)}_{:=b^*} \lambda + \underbrace{d_I d_V + (d_I -p_V) \beta_0 S_0}_{:= c^*},
\end{equation*}
 where $c^* < 0$, as $\frac{d_Id_V}{\beta_0S_0(p_V-d_I)} < 1$ according to the definition of $B_{\text{thres}}$. Then the eigenvalues are $\lambda_{3,4} = \frac{-b^* \pm \sqrt{b^{*2}-4a^*c^*}}{2a^*}$, with $a^* = 1$, and we get $\lambda_{3} < 0$ and $\lambda_{4} > 0$. Overall, this signifies that the disease-free steady state is unstable. Any initial amount of virus activates the immune response $B$, which then approaches $B_{\text{thres}}$.
Similarly, upon successful infection and recovery the post-infection steady state is given by 
\begin{center}
$(S^{**},I^{**},V^{**},B^{**}) = (S_0, 0, 0, B_{\text{thres}})$,
\end{center}
where the immune response reaches memory $B_{\text{thres}}$ countering reinfection of the same virus variant. 
The Jacobian at $(S^{**},I^{**},V^{**},B^{**})$ is given by
\begin{equation*}
J|_{(S^{**},I^{**},V^{**},B^{**})} = 
\begin{pmatrix}
-d_S & 0 & -\beta_0(1-B_{\text{thres}})S_0 & 0 \\
0 & -d_I & \beta_0(1-B_{\text{thres}})S_0 & 0 \\
0 & p_V & -d_V-\beta_0(1-B_{\text{thres}})S_0 & 0 \\
0 & 0 & p_B(1-B_{\text{thres}}) & -d_B B_{\text{thres}}
\end{pmatrix},
\end{equation*}
with 
\begin{align*}
\text{det}(J|_{(S^{**},I^{**},V^{**},B^{**})} - \lambda I) = &(-d_S - \lambda)(-d_I - \lambda) (-d_V -\beta_0 (1-B_{\text{thres}}) S_0 - \lambda) (-d_B B_{\text{thres}} - \lambda) \\
&-(-d_S - \lambda) p_V \beta_0 (1-B_{\text{thres}}) S_0 (-d_B B_{\text{thres}} - \lambda) \\
= &0.
\end{align*}
From this equation we get eigenvalues $\lambda_1 = -d_S < 0$ and $\lambda_2 = -d_B B_{\text{thres}} < 0$. For eigenvalues $\lambda_{3,4}$ we need to solve the remaining quadratic equation 
\begin{equation*}
    \lambda^2 + (d_V + \beta_0 (1-B_{\text{thres}}) S_0 + d_I) \lambda + \underbrace{d_I d_V + (d_I -p_V) (\beta_0  (1-B_{\text{thres}}) S_0)}_{:= c^{**}} = 0.
\end{equation*}
When inserting $B_{\text{thres}}$ from (\ref{eq:Bthres}), we see that $c^{**} = 0$ and hence, $\lambda_3 = -(d_V + \beta_0 (1-B_{\text{thres}}) S_0 + d_I) < 0$ and $\lambda_{4} = 0$. 
Thus, the investigation of stability of the steady state $X^{**} := (S_0, 0 ,0, B_{\text{thres}})$ 
is not determined by its linearization and requires higher order analysis.
According to the center manifold theorem 
(see e.g. Theorem 5.1 in \cite{Kuznetsov1998}), 
there is an one-dimensional invariant manifold (a curve) 
$C$ passing through $X^{**}$ tangent to the eigenvector $l_{(\lambda_4=0)}$, corresponding to the eigenvalue
$\lambda_4 = 0$. The  stability of the whole system is determined by its stability on the curve $C$. On this curve, the system can be either be instable, semi-stable (on one part of the curve only) or stable. The eigenvector $l_{(\lambda_4=0)}$ is given by 
\begin{displaymath}
       l_{(\lambda_4=0)} = \Big(-\frac{d_I d_V}{d_S}, \,d_V, \, p_V - d_I,\, 
              \frac{p_B(1- B_{\text{thres}})(p_V - d_I)}{d_B B_{\text{thres}}}\Big).   
\end{displaymath}
The analytical description of the center manifold and the investigation of its stability is non-trivial. Hence, we rely on the numerical results, which clearly show local stability of system (\ref{eq:ODEsystem}). In such a case, $C$ is a so-called ``slow manifold'': the convergence to the steady state
is not exponential but is at most so fast as $1/\sqrt{t}$ when time $t$ increases to infinity. This is also consistent
with the numerical results. There exists a third unique steady-state, however, not within the positive orthant, which is invariant. 
Since all eigenvalues are real and the trajectories of the system
converge locally to the steady state $X^{**}$, the latter is globally stable in the positive orthant.
Generally, this steady state can be approached asymptotically in one of two ways: either from above, when the short-term acute immune response is strong enough to drive $B(t) > B_{\text{thres}}$ in the initial phase of infection or from below. From the ODE system (\ref{eq:ODEsystem}) we see that $\frac{\partial B}{\partial t} = 0$ when
\begin{itemize}
\item[(i)] $(\bar{V}, \bar{B}) = (0,B_{\text{thres}})$ 
\item[(ii)] $(\bar{\bar{V}}, \bar{\bar{B}}) = \left( \frac{d_B(\bar{\bar{B}}-B_{\text{thres}})\bar{\bar{B}}}{p_B(1-\bar{\bar{B}})}, \bar{\bar{B}} \right)$. 
\end{itemize}
For the latter to make sense biologically $\bar{\bar{V}} \geq 0$, and hence it is required that $\bar{\bar{B}} \geq B_{\text{thres}}$. Whether the short-term acute immune response is strong enough to lead to $B \geq B_{\text{thres}}$ depends on the viral load $V$ and parameter $p_B$.  
The definitions of $B$ and $B_{\text{thres}}$ assure that the infection is fully and permanently cleared in the model, which is in line with current literature on SARS-CoV-2 for non-immunocompromised individuals. Without immune threshold at $B_{\text{thres}}$, the model would lead to a minimal but chronic infection with periodic spreading of the virus within the individual. 

\subsection*{Data and pre-processing of single-individual SARS-CoV-2 infection dynamics}
During fall 2020 and spring 2021, Ke et al. collected daily nasal samples for up to 14 days of all faculty, staff and students of the University of Illinois at Urbana-Champaign, who either (i) reported a positive quantitative reverse transcription polymerase chain reaction (RT-qPCR) result in the past 24 hours or (ii) were within five days of exposure to someone with a confirmed positive RT-qPCR result, while having tested negative for SARS-CoV-2 in the previous seven days \cite{Ke2022}. These criteria ensured a large-scale, high-frequency screening of early SARS-CoV-2 infection dynamics. In total, Ke et al. reported the cycle number (CN) values of nasal swab samples over time for 60 individuals. Due to very low or undetectable viral loads, we removed four out of the 60 individuals prior to the analysis, as also done by Ke et al. Time was reported relative to the day at which the maximal CN value was measured. 
The model is, however, initialized at infection. To compare measured and simulated CN values on the same time scale, we assumed viral load to peak six days post infection, as reported throughout literature for early SARS-CoV-2 variants \cite{Killingley2022, Puhach2023}. Hence, we shifted the relative time scale of the CN values measured by Ke et al. by six days and removed all measurements prior to the assumed day of infection.
Moreover, to directly compare CN values and simulated viral loads, we made use of the CN value-to-viral load calibration determined by Ke et al. and given by
\begin{equation}
\text{CN}  = -\frac{\text{log}_{10} V - 11.35}{0.25}.  
\end{equation}

\subsection*{Parameterization of the within-host SARS-CoV-2 model}
To describe the individual SARS-CoV-2 infection dynamics, we parameterized the model as shown in Table \ref{tab:nasal_saliva_par}.  

\begin{table}[h]
\begin{tabular}{p{0.25in} p{3.3in} p{1.35in} p{0.45in} }
 \hline
Variable & \vspace{0.1mm} & Initial value & Ref.\\
 \hline
$S$ & Number of susceptible cells & $S_0 = 8 \times 10^7$ &  \cite{Ke2022} \\
$I$ &  Number of infected cells & $I_0 = 1$ & - \\
$V$ & Number of measured virus & $V_0 = 0$ & -  \\
$B$ & Relative immune response & $B_0 = 0$ & - \\
 \hline
Rate constant & &  Value & Ref. \\
 \hline
$p_S$ & Susceptible cell production rate constant & $S_0 \times d_S$ /day&  - \\
$d_S$ &  Death rate constant of a susceptible cell & $0.091$ /day & \cite{Phan2023, Ruysseveldt2021} \\
$\beta_0$ & Infectivity constant & $4.92 \times 10^{-9}$ /day & \cite{Ke2022} \\
$d_I$ & Death rate constant of an infected cell & $2.45$ /day & \cite{Ke2022} \\
$p_V^*$ & Viral production rate constant of infected cell $\times$ sampled virus & $[10^2, 10^3]$ /day & - \\
$d_V$ & Viral clearance rate constant & $10$ /day & \cite{Ferretti2020,Szablewski2020} \\
$p_B^*$ & Activation rate constant of immune response &  $[10^{-10}, 10^{-4}]$ /day&  - \\
$d_B^*$ & Waning rate constant of immune response & $[10^{-4}, 1]$ /day &  -  \\
\end{tabular}
\caption{\textbf{Variables and rate constants of the model to describe the individual nasal viral load samples from Ke et al.}. Rate constants with $^*$ were estimated within the given upper and lower boundaries.}
\label{tab:nasal_saliva_par}
\end{table}

\noindent Initial values and the rate constants for five out of the eight parameters were taken from literature and assumed to be the same across individuals. The initial number of susceptible cells $S_0$ was derived by Ke et al. \cite{Ke2022}. Otherwise, we assumed the system to be initialized by a single successfully infected cell. Assuming that the system was in equilibrium during the initial disease-free state, we set the susceptible cell production rate constant $p_S$ to $S_0 \times d_S$, such that the in- and outflow of $S$ is the same. The rate constant values for the infection rate $\beta_0$ and the death rate of an infected cell $d_I$ were taken from Ke et al. and are the mean rate constants of their estimated individual-level rate constants \cite{Ke2022}. In the absence of an acute viral infection, the death rate constant of a susceptible cell $d_S$ is low, measured in the order of 0.02-0.03 /day \cite{Ruysseveldt2021}. However, this rate constant is estimated to increase during acute infection \cite{Phan2023}. For simplicity, we assumed a constant death rate constant between the lower and upper estimates. The value for the viral clearance rate constant $d_V$ was adopted from other studies \cite{Ferretti2020,Szablewski2020}. To account for the heterogeneity between individuals, we estimated the other three rate constants, namely (i) the viral production rate constant $p_V$, (ii) activation rate constant of the immune response  $p_B$, and (iii) waning rate constant of the immune response $d_B$ at an individual-specific level. The upper and lower boundaries of $p_V$ were determined by the range of estimated individual-specific rate constants from Ke et al. \cite{Ke2022}. The lower and upper boundaries of $p_B$ and $d_B$ are less interpretable and, hence, were assumed to cover a broad range of values.

\subsection*{Parameter estimation}
 Experimental data such as the measurements of CN values is noise corrupt. We took this measurement noise into account in the model, by assuming an additive Gaussian measurement noise distribution. The log-likelihood for the Gaussian noise model, for individual $i$, time point $k$ with measured CN value $\bar{y}_i^k$ is given by
 \begin{equation*}
     \text{logL}(\theta_i) = -\frac{1}{2} \sum_k \text{log}(2 \pi \sigma_i^2) + \frac{(\bar{y}_i^k - y(t_k,\theta_i))^2}{\sigma_i^2}.
 \end{equation*}
Hence, we also inferred an individual-specific noise parameter $\sigma_i$ per individual determining the spread of the Gaussian noise model. The lower and upper boundaries of $\sigma$ were set to $[10^{-2}, 10]$ according to the range of CN values measured. In total we estimated four individual-specific model parameters, $p_V$, $p_B$, $d_B$, and $\sigma$.
We performed multi-start maximum likelihood optimization of the negative log-likelihood in the log$_{10}$ parameter space for numerical reasons \cite{Hass2019}, initiating the optimization runs from 10 different Latin-hypercube-sampled starting points, maximizing over the CN values per individual. 

\subsection*{Numerical results}
The simulation study is based on the parameterization given in Table \ref{tab:nasal_saliva_par}. Based on the single-individual model fitting, we assumed five of the eight rate constants to be the same across all individuals, namely $p_S$, $d_S$, $\beta_0$, $d_I$, and $d_V$, while the remaining three rate constants, $p_V$, $p_B$, and $d_B$, were assumed to be individual-specific. These three rate constants were previously estimated for each of the 56 individuals measured by Ke et al. by using our model (Figure \ref{fig:sim_study}A). For the simulation study, we removed 14 out of the 56 individuals for which the estimated standardized rate constants of $p_V$, $p_B$, and $d_B$ were not within $\pm$2 $\times$ standard deviations (stds) of their respective standardized estimated distributions. For these individuals, the CN nasal swab samples have either not been collected during early infection, such that infection dynamics before peak viral load are not well captured by the model, or where the CN values did not include sufficient information to deduce the dynamics of long-term immune response level formation.  
We ensured that each of the three remaining standardized distributions were standard normally distributed by performing a one-sample Kolmogorov-Smirnov test employing the \textit{kstest} function of MATLAB \cite{kstest}. The function \textit{kstest} tests the null hypothesis that the parameter sample is drawn from a standard Gaussian distribution. As we required all three statistical tests to not reject the null hypothesis, we corrected for three-fold testing by applying the Bonferroni correction and adjusted the significance level of 0.05 to $\frac{0.05}{3} = 0.0167$ \cite{Armstrong2014}. The \textit{p}-values for the standardized estimated distributions of $p_V$, $p_B$, and $d_B$ are 0.48, 0.99, and 0.19, respectively. By drawing from this standardized multivariate-Gaussian distribution, we maintained the correlation between the parameters as estimated previously and assured that their values were within a plausible range. In total, we drew 50 rate constant triplets out of the resulting standardized multivariate-Gaussian distribution and used these re-transformed rate constants to simulate 50 nasal viral infection dynamics for up to 20 days post infection (Figure \ref{fig:sim_study}B).

\subsection*{Comparison of key features}
We compared five key features of the fitted and simulated infection dynamics, namely (i) the peak viral load, (ii) $B_{\text{thres}}$, (iii) the number of days from infection to peak viral load, (iv) the number of days from peak viral load to undetectable viral load, and (v) the number of days from infection to $B \geq B_{\text{thres}}$ (Figure \ref{fig:sim_study}B).
For the fits and simulations which did not reach undetectable CN values or for which $B \ngeq B_{\text{thres}}$ before 20 days post infection, we set these values to the maximum of 20 days. 
For each of the key features, we compared the resulting distributions from the fitted and simulated infection dynamics by performing a two-sample Kolmogorov-Smirnov test as provided by MATLAB function \textit{kstest2}. The function \textit{kstest2} tests for the null hypothesis that both parameter samples come from the same continuous distribution. As we tested for a single universal null hypothesis, where all five alternative null-hypotheses were required to not be rejected, we corrected for five-fold testing by applying the Bonferroni correction and adjusting the significance level of 0.05 to $\frac{0.05}{5} = 0.01$ \cite{Armstrong2014}. 

\subsection*{Reinfection}
We differentiated between two different variants employing distinct modes of reinfection: (i) increased infectivity or (ii) immune escape. Increased infectivity was introduced into the model by an increased infection rate constant $\tilde{\beta}_0 > \beta_0$ and ${B}_{\text{thres}}$ was modified to 
\begin{align}
\tilde{B}_{\text{thres}} = 1- \frac{d_I d_V}{\tilde{\beta}_{0} S_0 (p_V-d_I)},
\end{align}
according to its definition (equation (\ref{eq:Bthres}), Figure \ref{fig:reinfection}A). If $\tilde{\beta}_0 > \beta_{0}$, then $\tilde{B}_{\text{thres}} > B_{\text{thres}}$. Similarly, immune escape was introduced into the model by an initially decreased level of the variant-specific immune response, $\tilde{B}(0) < B(\tilde{t})$, where $\tilde{B}$ is the immune response of the new virus variant and $\tilde{t}$ is the time of reinfection (Figure \ref{fig:reinfection}B). For this virus variant, the immune threshold remains the same, such that $\tilde{B}_{\text{thres}} = B_{\text{thres}}$. For the simulations, we assumed individual $\#432192$ to have been exposed to one of the variants, leading to a single successfully infected cell at 90 days post primary infection. We then simulated the ODE system of (\ref{eq:ODEsystem}) for the new virus variant for another 60 days. The initial values of the number of susceptible cells $S$ and measured virus $V$, as well as the immune response $B$ for the simulations of reinfection, were taken from the long-term fit of individual $\#432192$ at 90 days since primary infection. We set the increased infection rate arbitrarily  to $\tilde{\beta}_0 = 2 \beta_0$ and immune escape to 50$\%$, such that $\tilde{B}(0) = 0.5 B(\tilde{t})$.

\subsection*{Implementation and code availability}
There is no original data underlying this work. Only previously published data was used for this study \cite{Ke2022}. The MATLAB code corresponding to this manuscript will be made available upon acceptance of the manuscript. The analysis was performed with MATLAB 2023a. 

\section*{Acknowledgements}
We would like to thank Dr. Walter Hugentobler for the insightful discussions regarding respiratory infections. The views expressed are purely those of the authors and may not in any circumstances be regarded as stating an official position of the European Commission.

\section*{Author Contributions}
Conceptualization, L.S., N.I.S.; 
Data curation, L.S.; 
Formal analysis, L.S., V.M.V.;
Funding acquisition, N/A;
Investigation, L.S.;
Methodology, L.S., N.I.S.;
Project administration, N.I.S.;
Resources, N.I.S.;
Software, L.S.;
Supervision, N.I.S.;
Validation, L.S.;
Visualization, L.S.;
Writing – original draft, L.S., V.M.V., N.I.S.;
Writing – review and editing, L.S., P.V.M., V.M.V., N.I.S.

\printbibliography 

\end{document}